\documentstyle[aps,preprint,epsfig]{revtex} \draft
\tightenlines

\def\LL{\left\langle}	% left angle bracket
\def\RR{\right\rangle}	% right angle bracket
\def\LP{\left(}		% left parenthesis
\def\RP{\right)}	% right parenthesis
\def\BE{\begin{equation}}
\def\EE{\end{equation}}
\def\BEA{\begin{eqnarray}}
\def\EEA{\end{eqnarray}}
\def\EL{\nonumber\\}

\begin{document}
%@twocolumnfalse\endcsname

\title{Measurement of hybrid content of heavy quarkonia using lattice NRQCD}

\author{ Tommy Burch, Kostas Orginos\footnote{Present address: RIKEN-BNL Research Center, Bldg 510a, Upton, NY 11973-5000}, and Doug~Toussaint }
\address{Department of Physics, University of Arizona, Tucson, AZ 85721, USA}

%Short form author list
%\author{
%Tommy Burch$\null^a$,
%Kostas Orginos$\null^a$
%and 
%Doug~Toussaint$\null^a$
%}
%\address{
%$\null^a$Department of Physics, University of Arizona, Tucson, AZ 85721, USA
%}
\date{\today}

\maketitle

\begin{abstract}\noindent
Using lowest-order lattice NRQCD to create heavy meson propagators and
applying the spin-dependent interaction,
$c_B^{} \frac{-g}{2m_q}\vec\sigma\cdot\vec{B}$, at varying intermediate time
slices, we compute the
off-diagonal matrix element of the Hamiltonian for the quarkonium-hybrid
two-state system. Thus far, we have results for one set of quenched
lattices with an interpolation in quark mass to match the
bottomonium spectrum. After diagonalization of the two-state Hamiltonian,
we find the ground state of the $\Upsilon$ to show a $0.0035(1)c_B^2$
(with $c_B^2 \sim 1.5-3.1$) probability admixture of hybrid,
$|b\bar{b}g\rangle$.
\end{abstract}
%]

% OUTLINE
% -hybrids and quarkonia
% -lattice nrqcd
% -our method (sigma.B at t')
% -results [<QQg|sigma.B|QQ> & sin(theta)]
% -future work (improvements in t & x, continuum extrapolation, phenomenology)
%
\narrowtext
\section{Introduction}

In quantum chromodynamics (QCD), it has been known for some time
that mesons ($q\bar{q}$) and baryons ($qqq$) are not the only composite
states for which gauge-invariant propagators may be constructed. Multi-quark
states ($qq\bar{q}\bar{q}$, $qqqq\bar{q}$, etc.) and states with
gluonic excitations, or hybrids ($q\bar{q}g$, $qqqg$, etc.), can also
form the necessary color singlet. It is possible for these states to have
quantum numbers ($J^{PC}$) different from those allowed for typical hadrons.
Such states are termed ``exotics'' and while they offer significant promise
for hybrid and multi-quark state detection, they will not be the subject of
our discussion here. Instead, we focus on ``non-exotic'' heavy hybrid mesons
($|Q\bar{Q}g\rangle$), those which have quarkonium-like quantum numbers. The
true, heavy meson ground state should thus be a mixture of heavy quarkonium
and hybrid:
\BE
|\Upsilon\rangle = A_0|Q\bar{Q}\rangle + A_1|Q\bar{Q}g\rangle + ....
\EE

\begin{figure}
\epsfxsize=3.5in
\epsfysize=1.5in
\epsfbox[0 0 580 270]{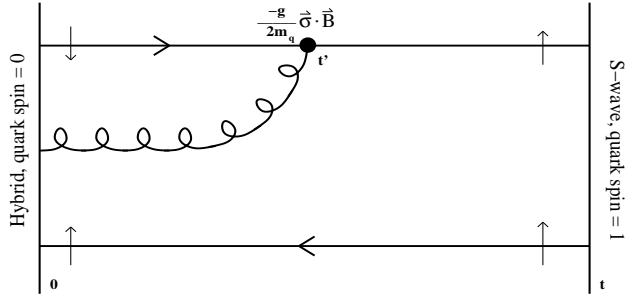}
\caption{
\label{HYBRIDMIXFIG}
Mixing of a $1^{--}$ hybrid with a $1^{--}$ S-wave via a single
application of the $\vec\sigma\cdot\vec B$ interaction.
}
\end{figure}

Heavy quarks within the bound state move with relatively small velocities
and thus we expect the rest mass of the quarks to dominate the energy
($K << m_q$). Expanding in
the quantity $\frac{1}{m_q}$, one finds a non-relativistic approximation
to the heavy-quark Hamiltonian (NRQCD) \cite{NRQCD_START}.
For simplicity, we keep only the lowest-order kinetic term of the NRQCD
Hamiltonian and the lowest-order spin-dependent term:
\BE
H = H_0 + \delta H = \frac{-\vec{D}^2}{2m_q} +
c_B^{}\frac{-g}{2m_q}\vec\sigma\cdot\vec{B} + ... ,
\EE
where $\vec{D}$ is the covariant derivative. Using only the first term in
this expansion, the Hamiltonian lacks a spin-flip interaction and there
is no mixing between the lowest-lying hybrid and quarkonium states.
For the $J^{PC}=1^{--}$ system, the total quark spin is 1 in the S-wave
state. Our operational definition
of the hybrid component of a $1^{--}$ meson is the component with total
quark spin of 0; the gluonic excitation (the $B_i$ in the meson
operator; see below) carries the angular momentum and ensures that the
color-octet channel of the quark/anti-quark state contributes to the
color-singlet meson propagator. Inclusion of the
spin-dependent term allows the mixing of these two states
(see Fig.~\ref{HYBRIDMIXFIG}).
Degeneracies are also lifted with the inclusion of this interaction:
e.g., the $0^{-+}/1^{--}$ mass difference is due to the hyperfine spin-spin
interaction and is quadratically dependent upon this term. Without the
$\vec\sigma\cdot\vec{B}$, these states are degenerate, as are the P-wave
states: $0^{++},1^{++},2^{++}$.

NRQCD has been used previously at this order, and beyond, to study heavy
hybrids \cite{NRQCD_HYBRIDS,UKQCD00}. Recently
Drummond {\em et al.}\ \cite{UKQCD00} reported seeing no quadratic dependence
of the $1^{--}$ hybrid mass (same source and sink operators) upon
the normalization of the $\vec\sigma\cdot\vec{B}$ term, although
the mass of the $0^{-+}$ hybrid did show effects.
The small, if any, dependence of the $1^{--}$ mass on $c_B$ can
be taken as evidence that the mixing with the S-wave $Q\bar Q$ state
is small.

We apply a different strategy, allowing us to measure the effect of the
spin-dependent interaction to first order. Rather than apply the
$\vec\sigma\cdot\vec{B}$ term at every
intermediate time slice in the lattice quark propagator, as has been the
usual practice with all terms in the NRQCD expansion
\cite{NRQCD_HYBRIDS,UKQCD00}, we restrict it to a particular intermediate
time slice, viewing this interaction as a ``perturbation'' at a specific
time in the lattice.
Thus, the unperturbed $Q\bar Q$ (quark-antiquark spin one) and unperturbed
$1^{--}$ hybrid (quark-antiquark spin zero) states propagate without mixing
except at the time slice where the perturbation is applied.
Then, using the non-exotic hybrid source
and the appropriate quarkonium sink (or vice versa), we extract the
off-diagonal matrix element of the $\vec\sigma\cdot\vec{B}$ interaction
for this two-state system. From this, we estimate the amount of
non-exotic hybrid admixture within the true ground-state using a simple
two-state mixing model. We expect that
such a result may be useful for studies of the creation and decay of heavy
quarkonia via color-octet channels (see, e.g., Ref.~\cite{TROTT94}) and
models of heavy-quark confinement \cite{SZCZEP97}.

\section{The Method}
We construct our lattice meson propagators using the NRQCD approach.
This effectively turns what would be a boundary value problem -
determining the relativistic quark propagators on a periodic lattice -
into an an initial value problem since, in the non-relativistic limit,
the quarks propagate only forward in time. Using this method, the evolution
of the quark propagator in the Euclidean time direction is given by:
\BEA
G(\vec{x}, t+a) = \LP 1 - \frac{aH_0}{2n}\RP^n U^\dagger_t(x)
\LP 1 - \frac{aH_0}{2n}\RP^n \EL
\cdot (1 - \delta_{t',t} a\delta H) G(\vec{x}, t) ,
\EEA
where $H_0$ and $\delta H$ are given above and $n$ is a parameter needed
for numerical stability \cite{LEPAGE92} ($n > \frac{3}{2m_qa}$; we use
$n = 2$). We also use plaquette tadpole improvement of the gauge links.
Note that we apply the interaction term, $\delta H$, at only a single
intermediate time step, $t'$.

We use an incoherent sum of point sources: at the source end, we start
with a given quark color and spin at all spatial points, without fixing the
gauge; at the sink end, we sum over all the contributions where the quark
and anti-quark are at the same spatial point. Since the lattices are not
gauge-fixed, we expect the contributions from sources with the quark and
anti-quark at different spatial points to average to zero.
We combine the quark and anti-quark sources (propagators) with the
appropriate spin matrices to construct the meson operators at the source
(sink) time slices. The meson operators we use are displayed in
Table~\ref{OPERATORS}.
The hybrid operators in Table~\ref{OPERATORS} involve the color magnetic
field, with $J^{PC}=1^{+-}$.  As can be seen from the quantum numbers,
this corresponds to a transverse electric gluon mode in a bag model approach.
The color magnetic field is calculated using the
``clover formulation'' (average of field from 4 plaquettes with corners at
these points):
\BEA
{\cal F}_{jk}(&x&) = \frac{1}{8}[U_j(x)U_k(x+\hat{j})U^\dagger_j(x+\hat{k})U^\dagger_k(x)\EL
&+& U_k(x)U^\dagger_j(x+\hat{k}-\hat{j})U^\dagger_k(x-\hat{j})U_j(x-\hat{j})\EL
&+& U^\dagger_j(x-\hat{j})U^\dagger_k(x-\hat{k}-\hat{j})U_j(x-\hat{k}-\hat{j})U_k(x-\hat{k})\EL
&+& U^\dagger_k(x-\hat{k})U_j(x-\hat{k})U_k(x-\hat{k}+\hat{j})U^\dagger_j(x)\EL
&-& \mbox{herm. conj.}]
\EEA
We use this form of the chromo-magnetic field (with the appropriate tadpole
factor: $1/u^4_0$) in the interaction term ($\delta H$).
The field used to make the hybrid sources and sinks is constructed in the
same fashion, with one exception: we use smeared links (sum of the simple
link and all 3-staples connecting neighboring lattice sites) in place of the
simple links in Eq. (4), the object being to improve the overlap with the
hybrid ground state.

Since we are working on lattices with a Euclidean metric (i.e.,
time is imaginary), the propagators should follow decaying exponentials.
The form we use to fit the meson correlators, $C(t)$, follows:
\BE
C(t) = A_0 e^{-m_0t} + A_1 e^{-m_1t}.
\EE
We include the second term to account for excited-state contributions.
The propagators were averaged over a set of quenched lattices with
Symanzik 1-loop improved gauge action.

To set the physical scale of our lattices, we use the S-P (spin-averaged)
mass difference for bottomonium, a quantity which is relatively insensitive
to the quark mass. We also create non-zero momentum operators for the
$1^{--}$ S-wave meson and determine its kinetic mass from the resulting
dispersion relation. An interpolation in $m_q$ is then performed to match
this kinetic mass with the experimentally determined value for the mass
of the $\Upsilon$. This provides us with an estimate of the bottom quark
mass, $m_b$.

For the ``mixed'' propagators (different source and sink operators), we
expect a propagator of the form
\BEA
C_{mix}(t) &=& A_{0,source}^{1/2}A_{0,sink}^{1/2}\LL 1^{--}(H)\left|c_B^{}
\frac{-g}{2m_q}\vec\sigma\cdot\vec{B}\right|1^{--}(S)\RR \EL
&\cdot& e^{-m_{0,source}t'} e^{-m_{0,sink}(t-t')} + ....
\EEA
Knowing the amplitudes and masses of the source and sink operators from
their ``unmixed'' propagators and fitting this propagator in the region
$t > t'$, we can extract the matrix element from the amplitude of
$C_{mix}(t)$ at different values of $t'$.
At sufficiently large values of $t'$, we expect less excited-state
contamination from the source operator and hope to find a plateau in the
value of the matrix element.

For these mixed propagators, we use the tree-level value for the
renormalization factor, $c_B^{}=1$, in the interaction term. The final result
for the matrix element, however, should contain the appropriate factor for
the given value of the lattice spacing. To address this, we choose a
non-perturbative approach. We perform additional spectrum runs, applying the
interaction term (with $c_B^{}=1$ and 2) at all intermediate time slices and
for both the quark and anti-quark. By interpolating in the resulting values
of the $0^{-+}/1^{--}$ mass difference to that of experiment (or, in the
present case of the $b\bar{b}$ system, potential model results), a value for
$c_B^2$ may be found.

\section{Results}
\begin{figure}
\epsfxsize=3.5in
\epsfysize=3.5in
\epsfbox[0 0 4096 4096]{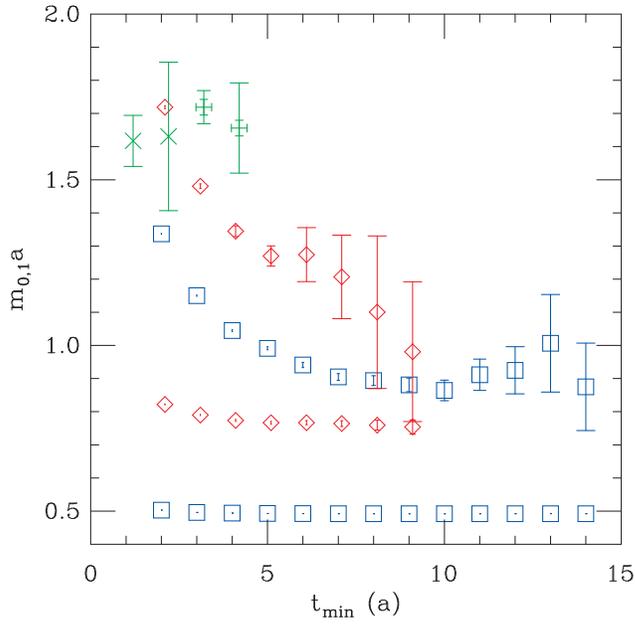}
\caption{
\label{FITMASSFIG}
Fit masses (above the zero point, $E_0$) vs. minimum time slice for the
$1^{--}$(S) (squares), $0^{++}$(P) (diamonds), and $1^{--}$ hybrid (cross:
two-mass fit; fancy plus: single-mass fit) with $m_qa = 2.5$. For the
$1^{--}$(S) and $0^{++}$(P), both ground state and first-excited state
masses are shown.
}
\end{figure}

The meson correlators were averaged over 165 quenched, $20^3 \times 64$,
$\beta=8.0$ lattices with Symanzik 1-loop improved gauge action (see
Ref.~\cite{MILC_SCALING} for more details on these lattices). Shown in
Fig.~\ref{FITMASSFIG} are the fit masses for the $1^{--}$(S) and $0^{++}$(P)
(ground and first-excited state), and the $1^{--}$ hybrid (ground state only).
The results of our chosen fits to the correlators appear in
Table~\ref{FITRESULTS}. Using the spin-averaged 1S-1P mass difference of 440
MeV for bottomonium, we find an inverse lattice spacing of
$a^{-1} = 1604(25)$ MeV [$a = 0.123(2)$ fm]. This differs from a
previous determination of the lattice spacing for this set of lattices using
the quantity $r_1/a$ from the static quark potential \cite{MILC_SCALING}:
$a^{-1} = 1449(4)$ MeV [$a = 0.1360(3)$ fm].

Using two values of the quark mass, $m_qa = 2.5$ and 2.8, we were able to
interpolate to a physical quark mass by fixing the kinetic S-wave mass to
that of experiment ($M_\Upsilon = 9.46$ GeV). We find a lattice-regularized
bottom quark mass of $m_b \approx 4.18$ GeV.

\begin{figure}
\epsfxsize=3.5in
\epsfysize=3.5in
\epsfbox[0 0 4096 4096]{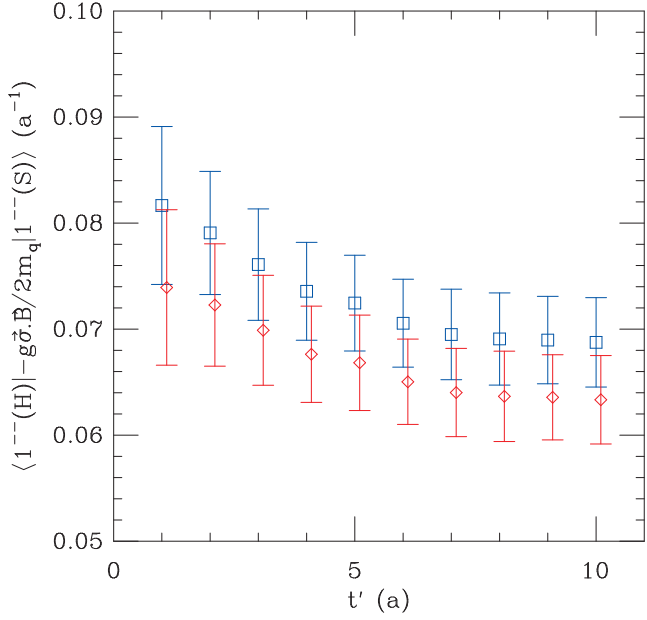}
\caption{
\label{SIGMABFIG}
Magnitude of the off-diagonal matrix element of the Hamiltonian (in lattice
units) for the $1^{--}$ S-wave (source) / hybrid (sink) two-state system vs.
the time slice, $t'$, at which the interaction ($\delta H$) is applied. The
squares are for $m_qa=2.5$, the diamonds for $m_qa=2.8$.
}
\end{figure}

\begin{figure}
\epsfxsize=3.5in
\epsfysize=3.5in
\epsfbox[0 0 4096 4096]{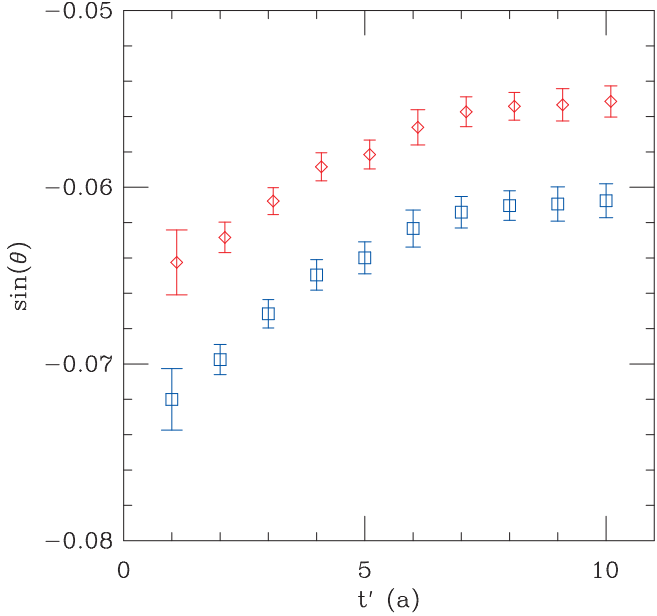}
\caption{
\label{SINSFIG}
Mixing angle, $\sin(\theta)$, vs. the time slice, $t'$, at which the
interaction term ($\delta H$) is applied. The squares are for $m_qa=2.5$, the
diamonds for $m_qa=2.8$.
}
\end{figure}

Fits to the ``mixed'' propagators were also performed and the resulting
values for the off-diagonal matrix element of the Hamiltonian appear in
Fig.~\ref{SIGMABFIG}, along with the associated jackknife errors. The
Hamiltonian was then diagonalized and the admixture of hybrid within the
true ground state calculated (see Fig.~\ref{SINSFIG}).
It may seem odd that the relative errors for $\sin(\theta)$ are much smaller
than those for the matrix element. However, the matrix element is quite
strongly correlated with the hybrid/S-wave mass difference and
since the mixing angle is roughly equivalent to the ratio of these two
quantities,
%-- $\sin(\theta) \approx \langle H|\vec{\sigma}\cdot\vec{B}|S\rangle
%/(m_S - m_H)$ --
the errors for $\sin(\theta)$ tend to be smaller. A plateau
is reached in these plots by $t'=8$.
Using the result at $t'=9$ ($\chi^2/d.o.f. < 1$) and interpolating in
the quark mass to $m_ba = 2.6$, we find
\BEA
|\Upsilon\rangle &=& \cos(\theta)|Q\bar{Q}\rangle +
\sin(\theta)|Q\bar{Q}g\rangle\EL
&=& 0.99826(6)|Q\bar{Q}\rangle - 0.059(1)|Q\bar{Q}g\rangle ,
\EEA
corresponding to 0.0035(1) probability admixture of hybrid in the $1^{--}$
bottomonium ground state.

While this result is clearly non-zero, there remain some
unresolved issues surrounding the actual numerical value.
For one thing, there is the question of the field normalization
(i.e., the value for $c_B^{}$). To get a handle
on this number, we performed additional spectrum runs with the interaction
``turned on'' for all times. The results for these propagator fits can be
found in Table~\ref{SBFITRESULTS}.
The resulting $0^{-+}/1^{--}$ mass differences ($\Delta M_{\Upsilon -\eta_b}$)
are found to be $\sim 20$ and $\sim 78$ MeV for $c_B^{}=1$ and 2, respectively.
While there is currently no experimental result for the $\eta_b$ mass, there
are some potential model calculations \cite{CHEN96,YNDURAIN01} which predict
this splitting to be in the $\sim 30-60$ MeV range. This would imply a value
of $c_B^2 \sim 1.5-3.1$, corresponding to the mixing angle values
$|\sin(\theta)| \sim 0.073-0.104$.
This range for our value of the mixing angle sits between two values found
previously with calculations based upon the MIT bag model \cite{BARNES79}:
$|\sin(\theta)| = 0.0427$ and 0.1503.
While a value of $c_B^{} \sim O(1)$ needed for consistency is encouraging,
a more precise numerical result for this parameter remains elusive, mainly
due to the fact that the actual $\Upsilon-\eta_b$ mass splitting is not well
known.

Comparison of our result for $\sin(\theta)$ with the hybrid mass results
of Ref.~\cite{UKQCD00} is not straight-forward. Whereas our result can
be related to s-channel diagrams where the valence gluon scatters
off the quark or anti-quark, there are other spin-dependent contributions
to the hybrid mass that cannot be taken into account with our first-order
result (see e.g., Ref.~\cite{BaClVi83}), including non-perturbative effects;
to include such contributions in the lattice simulation, one needs to apply
the $\vec{\sigma}\cdot\vec{B}$ interaction at all time slices. Since we do
have hybrid propagators with such a heavy-quark Hamiltonian, we attempted
to extract these hybrid masses for a more direct comparison with the
results of Ref.~\cite{UKQCD00}. However, as there is a significant
contribution to our hybrid propagators from intermediate mixing with the
corresponding S-wave states, these fits do not return reliable hybrid masses.
An effective mass plot would be better-suited for determining these hybrid
masses; however, our lattices lack the time resolution for finding the
necessary plateau at short times. Our time resolution is given by
$a^{-1}\approx 1600$ MeV, as compared with $a_t^{-1}\approx 4500$ MeV in
Ref.~\cite{UKQCD00}, where they find a plateau in their $1^{-+}$ hybrid
effective mass plot from $t\sim 4-10$. This suggests that we would need
finer resolution in the $1<t<4$ range to observe a similar plateau. It
may be for this reason that our former values (without
$\vec{\sigma}\cdot\vec{B}$) of the hybrid/S-wave mass splittings are quite
large: $\Delta M_{1H-1S} = 1.81(12)$ GeV. This is significantly above the
1.644(17) GeV result quoted in Ref.~\cite{UKQCD00} and the recent CP-PACS
result of 1.56(18) GeV \cite{CPPACS01}. However, if we use the lattice
spacing as determined by the static quark potential, $a^{-1} = 1.449(4)$
GeV and $m_ba \approx 2.9$, this splitting (which is relatively insensitive
to the quark mass) becomes 1.64(12) GeV. It should also be noted, however,
that our heavy quark action is simpler than those used by these groups in
that they include the first relativistic correction. The CP-PACS result
also includes two flavors of light dynamical quarks.

It would be useful to resolve these issues with a continuum extrapolation
as our results thus far are from quenched lattices with only a single value
of the coupling. It would also be useful to explore the effects of
dynamical quarks.

\section*{acknowledgements}

This work was supported by the U.S. Department of Energy under contract
DE FG03 95ER 40906. Computations were performed on the Nirvana cluster
at Los Alamos National Laboratory. We would like to thank Ted Barnes for
helpful suggestions.

%\begin{center}
\begin{table}
\caption{
\label{OPERATORS}
Meson operators.
}
\begin{tabular}{lc}
$J^{PC}$ & Operator \\ \hline
$0^{-+}$ S-wave ($\eta_b$) & $\bar{Q} Q$ \\
$1^{--}$ S-wave ($\Upsilon$) & $\bar{Q}\sigma_i Q$ \\
$0^{++}$ P-wave ($\chi_{b0}$) & $\bar{Q}\sigma_i D_i Q$ \\
$1^{++}$ P-wave ($\chi_{b1}$) & $\bar{Q}\varepsilon_{ijk}\sigma_j D_k Q$ \\
$2^{++}$ P-wave ($\chi_{b2}$) & $\bar{Q}(\sigma_i D_j + \sigma_j D_i - \frac{2}{3}\delta_{ij}\sigma_k D_k) Q$ \\
$0^{-+}$ hybrid & $\bar{Q} \sigma_i B_i Q$ \\
$1^{--}$ hybrid & $\bar{Q} B_i Q$ \\
\end{tabular}
\end{table}
%\end{center}

%\begin{center}
\begin{table}
\caption{
\label{FITRESULTS}
Fit results and resulting mass differences (with jackknife errors).
For each quantity, the first row is for $m_qa=2.5$, the second is for
$m_qa=2.8$.
}
\begin{tabular}{clll}
Propagator & Fit range & $m_0a$ & $\chi^2/d.o.f.$ \\ \hline
$1^{--}$(S) & 9-25 & 0.4920(2) & 13/13 \\
$''$ & 9-25 & 0.4754(2) & 16/13 \\
$0^{++}$(P) & 5-21 & 0.767(4) & 5.4/13 \\
$''$ & 5-21 & 0.749(4) & 5.8/13 \\
$1^{--}$(H) & 1-6 & 1.62(8) & 0.16/2 \\
$''$ & 1-6 & 1.62(8) & 0.01/2 \\ \hline
Quantity & & Mass ($a^{-1}$) & Mass (MeV) \\ \hline
$\Delta M_{1P-1S}$ & - & 0.275(5) & 440(fixed) \\
$''$ & - & 0.273(4) & 440(fixed) \\
%$\Delta M_{2S-1S}$ & - & 0.39(2) & 620(30) \\
%$''$ & - & 0.38(2) & 610(30) \\
$\Delta M_{1H-1S}$ & - & 1.13(8) & 1800(120) \\
$''$ & - & 1.14(8) & 1840(130) \\
$M^{kinetic}_{1^{--}(S)}$ & - & 5.64(6) & 9030(100) \\
$''$ & - & 6.36(6) & 10230(100) \\
\end{tabular}
\end{table}
%\end{center}

%\begin{center}
\begin{table}
\caption{
\label{SBFITRESULTS}
Fit results and resulting mass differences (with jackknife errors)
with the interaction term present at all intermediate time slices.
For each quantity, the first row is for $c_B^{}=1$, the second is for
$c_B^{}=2$ ($m_qa=2.5$).
%$^*$Second-lightest mass in a three-mass fit
%(the lightest masses are consistent with the corresponding lightest
%S-wave masses).
}
\begin{tabular}{ccll}
Propagator & Fit range & $m_0a$ & $\chi^2/d.o.f.$ \\ \hline
$0^{-+}$(S) & 8-24 & 0.4736(2) & 20/13 \\
$''$ & 8-24 & 0.4222(2) & 14/13 \\
$1^{--}$(S) & 8-24 & 0.4859(2) & 12/13 \\
$''$ & 8-24 & 0.4680(3) & 7.4/13 \\
$0^{++}$(P) & 5-21 & 0.764(5) & 7.8/13 \\
$''$ & 5-21 & 0.737(5) & 10/13 \\
$1^{++}$(P) & 5-21 & 0.762(6) & 5.8/13 \\
$''$ & 5-21 & 0.737(8) & 13/13 \\
$2^{++}$(P) & 5-21 & 0.753(6) & 2.1/13 \\
$''$ & 5-21 & 0.721(7) & 6.0/13 \\ \hline
%$0^{-+}$(H) & 1-16 & 1.77(7)$^*$ & 12/10 \\
%$''$ & 1-16 & 1.41(4)$^*$ & 14/10 \\
%$1^{--}$(H) & 1-16 & 1.99(14)$^*$ & 9.6/10 \\
%$''$ & 1-16 & 1.74(10)$^*$ & 9.0/10 \\ \hline
Quantity & & Mass ($a^{-1}$) & Mass (MeV) \\ \hline
$\Delta M_{\bar{\chi}-\Upsilon}$ & - & 0.271(5) & 440(fixed) \\
$''$ & - & 0.260(7) & 440(fixed) \\
$\Delta M_{\Upsilon-\eta_b}$ & - & 0.01232(8) & 20.0(4) \\
$''$ & - & 0.04583(24) & 78(2) \\
%$\Delta M_{\eta_b(H)-\Upsilon}$ & - & 1.29(7) & 2090(110) \\
%$''$ & - & 0.94(4) & 1590(70) \\
%$\Delta M_{\Upsilon(H)-\Upsilon}$ & - & 1.51(14) & 2450(230) \\
%$''$ & - & 1.27(10) & 2150(170) \\
\end{tabular}
\end{table}
%\end{center}

%\widetext
%\narrowtext

%\widetext
%\narrowtext

%\widetext
%\narrowtext

%\widetext
%\narrowtext


\begin{thebibliography}{99}

\bibitem{NRQCD_START}
W. E. Caswell and G. P. Lepage, Phys. Lett. B {\bf 167}, 437 (1986);
E. Eichten, Nucl. Phys. Proc. Suppl. {\bf B4}, 170 (1988);
G. P. Lepage and B. A. Thacker, Nucl. Phys. Proc. Suppl. {\bf B4}, 199 (1988);
G. P. Lepage and B. A. Thacker, Phys. Rev. D {\bf 43}, 196 (1991).

\bibitem{NRQCD_HYBRIDS}
T. Manke {\em et al.}, Phys. Rev. D {\bf 57}, 3829 (1998);
T. Manke {\em et al.}, Phys. Rev. Lett. {\bf 82}, 4396 (1999);
A. Ali Khan {\em et al.}, Nucl. Phys. Proc. Suppl. {\bf B83}, 319 (2000);
K. J. Juge, J. Kuti, and C. J. Morningstar, Nucl. Phys. Proc. Suppl. {\bf B83}, 304 (2000).

\bibitem{UKQCD00}
I. T. Drummond {\em et al.}, Phys. Lett. B {\bf 478}, 151 (2000).

\bibitem{TROTT94}
H. D. Trottier, Phys. Lett. B {\bf 320}, 145 (1994).

\bibitem{SZCZEP97}
A. P. Szczepaniak and E. S. Swanson, Phys. Rev. D {\bf 55}, 3987 (1997).

\bibitem{LEPAGE92}
G. P. Lepage {\em et al.}, Phys. Rev. D {\bf 46}, 4052 (1992).

\bibitem{MILC_SCALING}
C. Bernard {\em et al.}, Phys. Rev. D {\bf 62}, 034503 (2000).

\bibitem{CHEN96}
Y.-Q. Chen and R. J. Oakes, Phys. Rev. D {\bf 53}, 5051 (1996).

\bibitem{YNDURAIN01}
F. J. Yndurain, Nucl. Phys. Proc. Suppl. {\bf 93}, 196 (2001).

\bibitem{BARNES79}
T. Barnes, Nucl. Phys. {\bf B158}, 171 (1979).

\bibitem{BaClVi83}
T. Barnes, F. E. Close, and F. de Viron, Nucl. Phys. {\bf B224}, 241 (1983).

\bibitem{CPPACS01}
T. Manke {\em et al.}, hep-lat/0103015.

\end{thebibliography}
\end{document}